# How Does library Migration Impact Software Quality and Comprehension? An Empirical Study

Hussein Alrubaye[1], Deema Alshoaibi[1], Eman Alomar[1], Mohamed Wiem Mkaouer[1], and Ali Ouni[2]

[1] Rochester Institute of Technology, Rochester, New York, United States
{hat6622, da3352, eaa6167, mwmvse}@rit.edu
[2] Ecole de Technologie Superieure, University of Quebec, QC, Canada
ali.ouni@etsmtl.ca

**Abstract.** The process of migration between different third-party software libraries, while being an typical library reuse practice, is complex, time consuming and error-prone. Typically, during a library migration process, developers opt to replace methods from a retired library with other methods from a new library without altering the software behavior. However, the extent to which the process of migrating to new libraries will be rewarded with improved software quality is still unknown. In this paper, our goal is to study the impact of library API migration on software quality. We conducted a large-scale empirical study on 9 popular API migrations, collected from a corpus of 57,447 open-source Java projects. We computed the values of commonly-used software quality metrics before and after a migration occurs. The statistical analysis of the obtained results provides evidence that library migrations are likely to improve different software quality attributes including significantly *reduced coupling, increased cohesion, and improved code readability*. Furthermore, we released an online portal that helps software developers to understand the impact of a library migration on software quality and recommend migration examples that adopt best design and implementation practices to improve software quality. Finally, we provide the software engineering community with a large scale dataset to foster research in software library migration.

## 1 Introduction

Software maintenance activities consume up to 70% of the total life-cycle cost of a typical software product [9]. To cope with the expense software evolution, software reuse, through third-party libraries and APIs, has become the backbone of modern software development. However, just like any traditional code, libraries and APIs undergo maintenance and evolution. In order to keep most up to date libraries, developers need to periodically perform *third-party library migration* [26,27]. In practice, library migration can be seen as the process of replacing a library with a different one, while preserving the same program behavior. The library migration process tends to be a manual, error-prone, and time-consuming process [14,28,3,17,7,4]. Hence, developers have to explore and understand the new library's API, its associated documentation, and its usage scenarios in order to find the right API method(s) to replace in the current implementation belonging to the retired library's API. As a consequence, developers often spend considerable time to verify that the newly adopted features do not introduce any regression. Indeed, previous studies have shown that developers typically spend up to 42 days to migrate between libraries [6].

Unlike library upgrades, library migrations typically require more fine-grained code changes and refactorings, *e.g.*, changing types of variables and parameters, renaming attributes and methods, etc, since developers need to accommodate the syntactic and semantic mismatch between the added and removed methods [26]. These refactoring changes may account for the overhead needed to fulfill the migration and adjust the existing software design to the newly introduced methods. Typically, API migrations introduce a set of methods and objects with different lexical and naming conventions, which have to be integrated into the existing codebase terminology. That is, developers may refactor their code along with the migration to contextualize the new library methods. These unintended refactoring operations have an impact on software design metrics (*e.g.*, cohesion, coupling, etc.) [13] beside the changes in terminology and renaming activities that affects code readability as well [22,1,10].

Various studies have focused on analyzing the impact of API evolution on software quality in terms of change and bug-proneness [21,19,16], software usability and rating [18,8]. Other studies focused on estimating the impact of API documentation on the library adoption and usability which has been investigated in the literature [15]. Moreover, recent studies attempted to identify traces of manually performed library migrations. They provide the community of a set of real-world migrations between popular Java libraries, in various open source projects [27,28,6].

Existing studies reveal the importance of taking into account the software design characteristics when performing the migration to reduce maintenance costs. However, there is little knowledge on the impact of API migration, and its related refactoring changes, on the quality of software's design as well as code comprehension and readability. Indeed, as software systems evolve rapidly, there is a need for appropriate tools, reliable, and efficient techniques to support developers in replacing their deprecated library APIs with up-to-date ones, and maintaining/improving the quality of their software design.

To address the above-mentioned issues, we conducted a large-scale empirical study to assess the impact of library migration on both software design quality and code comprehension. We considered an existing dataset of 9 popular migrations between Java libraries, mined from 57,447 open-source Java projects [6]. Afterward, we shortlisted all commits containing traces of method swaps, as part of migrations under the study. We refined our dataset by untangling each commit to identify the specific code elements involved in the migration using program analysis. Then, for the selected code elements, we calculated the values of their corresponding design and readability metrics, before and after the migration. Finally, we statistically compared the variation of these values, to an-alyze whether the migration had a significant, positive, or negative impact on design quality and readability. We finally associated a ranking score, to each migration trace, according to the extent to which it was able to improve the design and readability of the existing code. We survey 10 senior developers to assess the usefulness of the ranking score in providing relevant migration code examples.

Our study is driven by the following research questions:

**RQ1. (Design Improvement) What is the impact of library migration on the quality of software design?**
To answer this research question, we assess the impact of library migration on software design quality in terms of complexity, coupling and cohesion, widely popular structural metrics [24], and previously used metrics in similar empirical studies [12,20]. For each analyzed source file in the dataset (that we detail later in the next subsection), we measured the value of its coupling and cohesion before and after the migration. As we aggregated all values before and after the migration, we observed the variation in the aggregated values to investigate whether the migration had a positive or negative impact on design quality.

**RQ2. (Code Readability) Does migration improve the code readability?**
Similarly to RQ1, we consider popular state-of-the-art readability tools and metrics [10,23]. For each metric, we measure its pair values in the dataset files, before and after the migration, and then we analyze the values for statistical significance.

**RQ3. (Quality Recommendation) Can we leverage design and readability metrics to recommend better code examples for migration?**
Since there are multiple code fragments, belonging to various projects and containing the same mappings, we design a recommendation-based ranking method that aggregates various quality metrics. Our method ranks the collected code fragments based on the extent to which they preserve the design coherence and improve the code comprehension. We then perform a qualitative study with 10 senior developers to evaluate the usefulness of our recommendation-based ranking method.

The paper's key findings show a positive variation of structural and readability metrics, *i.e.*, developers do pay attention to design and readability when performing the migration process. Moreover, results show that code fragments with higher ranking score were also voted by the majority of developers, as good examples of migrations. This study makes the following contributions:

1. We release an online portal[3] that showcases real-world migration fragments, with their corresponding positive or negative impact on coupling, cohesion, and readability.
2. We propose a ranking score, that we label *Migration Quality Score* (MQS), which ensures better API reuse by recommending migration examples that ensure better software quality and comprehension.
3. We survey with senior software engineers at an outstanding company to evaluate MQS's ability to recommend high-quality migration examples for 9 popular migrations. Findings show that MQS effectively recommends high-quality migration examples that facilitates API reusability.

---
[3] http://migrationlab.net/index.php?cf=icsr2020

## 2 Background and Terminology

This section provides basic terminology and concepts about library migration and design metrics used in our study.

### 2.1 Library migration

When a software development team made the decision of replacing the current libraries used by the software, they have to specify the **migration rules**. A migration rule is denoted by a pair of a *source* (retired) library and a *target* (replacing) library, *i.e.*, *source* → *target*. For example, *easymock* → *mockito* represent a migration rule where the library easymock is migrated to the new library mockito. Migration rules are not enough to start the migration process. Developers should define the mappings between methods. **Method mapping** is the process of replacing at least one method from the source library by one or multiple methods belonging to the target library.

### 2.2 Migration Example

We showcase, in Figure 1, a real-world example of a method-level migration as part of replacing the *json* library with the *gson* library[4]. The method *put(key, value)* has been replaced with two methods, namely *addProperty(key, value)*, and *Gson().toJson(value)*. To have valid input for *addProperty* method, the *Map* object needs to be converted into a *json* object, so another converting method was added. We call this type of change as *Migration fragment* where a block of code changes has methods from removed/added libraries.

```
public void addKeyValues(String key, Map value) {
    checkIfKeyDescriptionExist(key);
-   keyValues.put(key, value);
+   keyValues.addProperty(key, new Gson().toJson(value));
}
```

Fig. 1: Sample of migration between *json* and *gson*.

---

[4] http://migrationlab.net/redirect.php?cf=icpc2019&p=1

### 2.3 Software quality attributes

Object oriented (OO) software quality attributes reflects the quality change of a refactoring operation. The wide used attributes for software structure design and size are coupling, cohesion and complexity. Coupling measures the level of relationship between modules [25]. While designing the software, low coupling is desirable (*i.e.,* less dependency between modules). Coupling Between Objects (CBO) is a metric for measuring coupling between code objects. The higher the CBO, the higher the class coupling. Cohesion measures the level of relationship within module [25]. While designing the software, high cohesion is desirable (*i.e.,* strong interaction between code elements in a module) since this target helps in fostering code maintainability. Cohesion of Methods (LCOM) metric is used to assess the cohesion of classes. Normalized LCOM metric has been widely recognized in the literature [20,12] as being the alternative to the original LCOM, as the latter addresses its main limitations (misperception of getters and setters, etc.). The lower the LCOM, the higher the class cohesion. Complexity of software indicates effort and time required to maintain the software. Complex software costs more during maintenance and refactoring. Five complexity and volume metrics are used to compute this quality attribute, namely, the Cyclomatic Complexity (CycC), the Line of Code (LOC), the Line with Comments (CLOC), the Ratio of Comment Lines to Code Lines, and the Number of Blank Lines. Normally, higher values of these metrics indicate a higher value of class complexity [13].

Code readability impacts further code changes conducted by different team members than the original developer or even for the same developer but after a while. Source code readability is one of the important aspects of software engineering. Line length and number of comments are the basic readability metrics obtained from code static analysis. Buse and Weimer [10] derived a relationship between code metrics and human readability notation. Scalabrino et al. [23] extracted code textual features from source code lexicon analysis. The validity and usability of those readability metrics were tested by humans and show high correlation between human conception of code readability and metrics values.

## 3 Empirical Study Setup

### 3.1 Approach

Figure 2 provides an overview of our study workflow. To measure the impact of library migration on software quality attributes, we need to analysis the source code before and after library migration has happened. To do so, we used MigrationMiner [5], a command-line based tool used to detect library migration at the method level. Given 57,447 GitHub Java projects which provided by Allamanis et al. [2] as input to MigrationMiner [5], The tool detects 8,938 *migration commits* where a developer migrates the project's source code from using library A to library B (ex *easymock* → *mockito*). To analyze the impact of library migration on code quality, we run *Understand*, *readability.jar* [10,23] on migration commit( Commit N) and a commit before migration( Commit N-1).

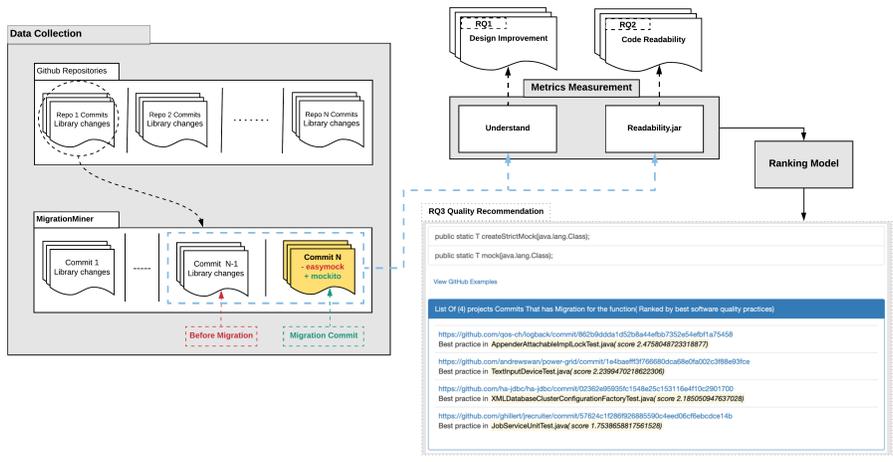

Fig. 2: Experimental Design Overview.

Each *migration commit* contains at least one or multiple mappings, *i.e.*, fragments of code containing one or multiple removed methods, being replaced with one or multiple added methods, along with other code changes that may or may not be related to the migration. Since any code change, not related to migration represents a noise for this study, we only consider files containing migration fragments in each *migration commit*. We notice that some migrations are *instant i.e.*, all method replacements are located in the same commit, but in multiple source files, and some migrations are *delayed*, *i.e.,* method replacements are scattered across multiple commits.

The data collection process has analyzed commits belonging to a diverse set of 57,447 projects. We have identified 36,023 classes, each contains at least one mapping. We also enumerated 9,380 unique mappings, already showcased in the dataset's website[5]. Also, we provide our collected data for replication and extension online.

After applying these tools on all predefined *mappings commits*, before and after the migration, we generate a dataset that contains, for each commit, its associated code fragments, structural and readability metrics pairs of values. We then use this dataset as a base of examples that we rank according to how much they improve quality and comprehension. We detail our proposed ranking model in the following Section 3.2.

---

[5] http://migrationlab.net/index.php?cf=icsr2020

| Property | # of instances |
|---|---:|
| # unique migrations | 9 |
| # projects | 57,447 |
| # commits have Migration | 393 |
| # Classes involved in migration | 36,023 |

Table 1: Dataset overview.

**Collecting Quality metrics measurements** To collect the design metrics, we use, Scitools Understand, a static analysis framework that captures a variety of structural metrics, across languages such as C++ and Java. Based on the computed metrics values, we can calculate the effect of migration-related changes on the system design. In particular, we analyze the following size and structure metrics: Coupling Between Objects (CBO), normalized Lack of Cohesion (LCOM), and Cyclomatic complexity(CycC).

Since each source file may contain multiple migration fragments, and since we only care about these specific files, we calculate metrics only for these fragments and then we average them to construct one value per file. In other terms, each data point in our analysis is a file with an average metric value.

Code readability during the migration process was measured by two state-of-the-art metrics proposed by Buse and Weimer [10], and Scalabrino et al. [23]. On the one hand, Buse and Weimer's Readability metric (BWR) combines the source code size characteristics to approximate its readability. On the other hand, Scalabrino et al.'s Readability metric (SR) does not only look at the structural characteristics of code, and adds another lexical dimension, in which it considers more linguistic properties such as comments consistency with the source code and its coherence etc. Both metrics generate a score that, the higher it is, the better is the readability of the code.

Similarly to structural metrics, each data point in our analysis represents an average readability score per source file.

### 3.2 Ranking Model

The migration dataset [6] contains, for each migration rule, *e.g.*, *easymock* to *mockito*, several commits, extracted from various projects, containing similar mappings. Therefore, for the same mapping, there are various real-world examples of how a deprecated method has been replaced with one or multiple replacing methods. Although these examples exhibit similar sets of removed/added methods, they differ in their overhead in the software design, since the migration process is subjective [27,28,7], and developers may perform different types of code changes to perform the same type of migration. Moreover, as maintaining a good quality of the source code, in terms of design and readability, is critical for code longevity, our aim is to favor the recommendation of source code migration examples that correctly execute the migration while also maintaining, or improving the current client code quality. To do so, we simply leverage the existing software quality attributes metrics, previously explored in the background section, and combine them into an overall *Mapping Quality Score* (MQS). For each given migration in the dataset, we loop through all its mappings, for each mapping, we locate all its instances in the course code (*inst*). Then, for each instance, we calculate its MQS, and finally, we rank them on a descendent order, to favor examples with the highest quality improvement. Formally, we calculate the MQS as follows:

$$MQS(inst) = \sum_{i=1}^{Metrics} W_i^{MQS} * \varphi_i(inst) \qquad (1)$$

where MQS represents the weighted sum of all values $\varphi_i$, and *i* varies according *Metrics*, the number of metrics used. The term *inst* denotes code instances to be ranked for a given mapping.

Since the combined metrics do not belong to the same scale, we normalize them using *min-max normalizer* that linearly rescales every metric value to the [0,1] interval. Rescaling in the [0,1] interval is done by shifting the values of each feature *x* so that the minimal value is 0, and then dividing by the new maximal value (which is the difference between the original maximal $max(x)$ and minimal $min(x)$ values).

Moreover, since not all metrics are to be maximized, we transform all of them to be minimized using the duality principle. For example, since the lower are the values of coupling, the better they are, we maximize the complement of the normalized value of coupling, *i.e.*, $\varphi_{CBO} = (1\text{-}z(CBO(src)))$, where *z* returns the min-max normalized value.

As an illustrative example, we observe in Figure 2 that for a given mapping between `createStrictMock`, belonging to the removed library *easymock*, and `mock`, belonging to *mockito*, 4 instances are being shown and recommended as migration examples. Note that each example contains a link to the actual location of the code on GitHub. The examples have been ranked according to their MQS. For instance, the first example has the highest MQS of 2.475, while the second example has an MQS of 2.239.

Note that the normalization was restricted to the MQS calculation, we still use the actual *raw* values of the metrics for the results, which are detailed in the following sections. Also note that we weights for the actual MQS score are by default equal to 1 i.e., for this study, we consider all metrics to be equally important, and thus, this can be improved, if any metric has been found to be more influential than others in this context of API migration.

This section details the results of our empirical setup to answer the research questions.

### 3.3 RQ1. (Design Improvement) What is the impact of library migration on the quality of software design?

Figure 3 outlines the box plots of the values, for each of the structural metrics, calculated before and after the migration. To better understand the statistical significance of the observed results, we setup our statistical analysis as follows: for each metric, we cluster its values according to whether it was measured before or after the migration. We apply this to each code fragment. As a result, we create two groups of equal size, each containing measurements of the same metric before and after the migration. Then, we use the Wilcoxon signed rank test, since these groups are dependent (measurement on the same code fragments), to evaluate the significance of the difference between the values, in terms of their mean.

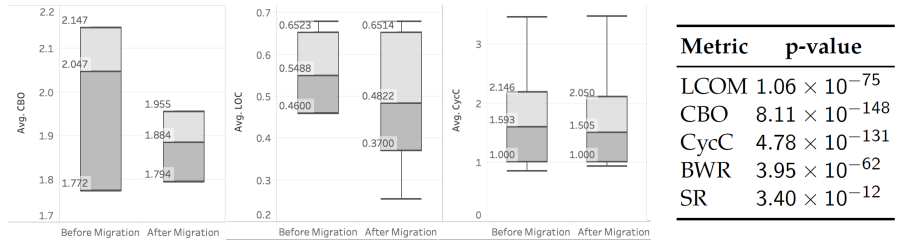

Fig. 3: Box plots of CBO, LCOM, and average CycC values, extracted from migrated code fragments, before and after the migration (lower values are better). Statistical significance of samples difference, before and after API migration, for each of the considered metrics.

Our Null hypothesis indicates no variation in the metric values of pre- and post-migrated code elements. In contrast, the alternative hypothesis advocates for a variation in the metric values. In this research question, a decrease in the mean values is considered desirable (*i.e.*, an improvement in design quality). Additionally, the variation between values of both sets is considered significant if its corresponding p-value is less than 0.05 (a confidence level of 95%). We deploy the same statistical analysis for RQ2 as well, but with a difference in the interpretation, since for readability metrics, an increase in mean value is considered desirable.

As can be seen in Figure 3, for the coupling between objects metric (CBO), we clearly notice a general trend of values being significantly decreased, just after the migration. The mean CBO value has decreased from 2.047 to 1.884 ($p-value < 0.05$), and the upper quartile has become significantly lower while decreasing from 2.147 to 1.955. Interestingly, we also observe from the figure a similar trend for the Lack of Cohesion of Methods metric (LCOM), since its mean value has gone from 0.548 to 0.482 ($p-value < 0.05$). We also notice a drop in the lower quartile, going from 0.460 to 0.370.

As for the average Cyclomatic complexity, there is a slight decrease in the upper quartile, varying from 2.146 to 2.050, but the mean value has decreased from 1.593 to 1.505 ($p-value < 0.05$).

To better understand the observed results, we manually analyze few random instances. Figure 4 illustrates a code fragment example of such migra-

```
...der/src/test/java/org/jboss/aerogear/unifiedpush/test/archive/UnifiedPushArchiveImpl.java
@@ -93,8 +93,7 @@ public UnifiedPushArchive withApi() {

     @Override
     public UnifiedPushArchive withUtils() {
-        return addPackage(org.jboss.aerogear.unifiedpush.utils.AeroGearLogger.class.getPackage())
-                .addClasses(ConfigurationUtils.class);
+        return addClasses(ConfigurationUtils.class);
     }
```

Fig. 4: Illustrative example of a code migration from *log4j* to *slf4j*, with a positive impact on coupling.

tions, extracted from Github [6]. In this fragment, the methods `addPackage` with `addClasses`, belonging to the library *log4j*, is being replaced with the method `addClasses`, from *slf4j*. We can observe the difference in the used parameters between the replaced and replacing methods. More precisely, `addPackage` with `addClasses` have a CBO of 4, while `addClasses` only have a CBO of 3, which did improve the overall CBO of all methods by adopting this newly deployed method.

Another interesting example in Figure 5[7] shows how the newly introduced object `DefaultHttpClient` does not rely on any parameter, unlike the retired object `HttpCLient` whose constructor is initialized with `connectionManager`. Therefore, the new object is more cohesive and it reduces class lack of cohesion.

> **Summary for RQ1.** Our empirical analysis has shown that APIs migration exhibit a positive impact on the software's design quality, in terms of complexity, coupling and cohesion.

### 3.4 RQ2. (Code Readability) Does migration improve the code readability?

Figure 6a outlines the boxplots of the values, for each of the readability metrics, calculated before and after each API migration.

For the BWR [10] metric, we observe an improvement in its values. In particular, the mean BWR [10] value has increased from 0.474 to 0.482 ($p-value < 0.05$). Similarly, the lower and the upper quartiles have slightly increased respectively from 0.316 to 0.329, and 0.579 to 0.587. As for the second readability metric, namely SR [23], the improvement is more significant since its mean value exhibits an increase from 0.568 to 0.603 ($p-value < 0.05$). The increase is also seen in the lower quartile, going from 0.461 to 0.484, whereas the upper quartile exhibits a slight decrease from 0.709 to 0.706.

---

[6] https://github.com/aerogear/aerogear-unifiedpush-server/commit/4861157566723bc3179b69d0755e5bf5460d9729

[7] https://github.com/anthonydahanne/ReGalAndroid/commit/6410cc8a12246745b19a102da5dd2c92d326b9f9

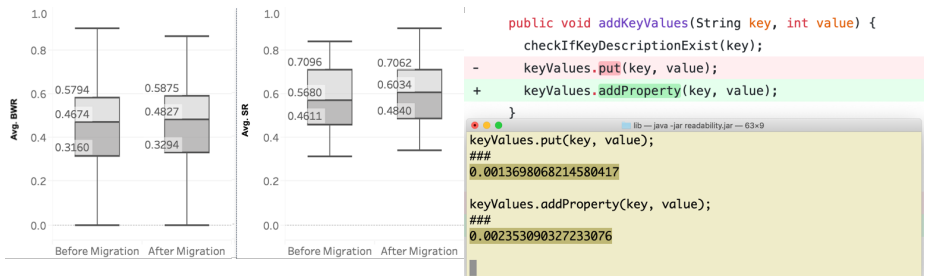

Fig. 5: Illustrative example of a code migration from *async-http-client* to *httpclient*, with a positive impact on cohesion.

If we take deeper look into the code example[8], illustrated in Figure 6b, we notice that the developer just moved from using the method `put`, from *json* to the method `addProperty`, from *gson*. Note that the developer did not perform any additional activities; however the BWR [10] improved from 0.0013 to 0.0023 since the method name *addProperty* has better readability score than *put*, as shown in the console output of BWR [10] in Figure 6b.

(a) Box plots of BWR and SR values, extracted from migrated code fragments, before and after the migration (higher values are better).

(b) Illustrative example of a code migration from *json* to *gson*, with a positive impact on readability.

Fig. 6: Impact of API Migration on Code Readability Metrics.

---

[8] https://github.com/groupon/Selenium-Grid-Extras/commit/4d9bada8aeab5b09e7a27926fc9ecab8bb5a1b51

> **Summary for RQ2.** API migrations do improve code readability, as both BWR [10] and SR [23] readability metrics experience a significant increase when comparing code fragments before and after the migration.

### 3.5 RQ3. (Quality Recommendation) Can we leverage design and readability metrics to recommend better code examples of migration?

To evaluate our ranking model based on the structural and readability metrics, we conducted a qualitative analysis with 10 senior developers from an outstanding software development company. All the participants volunteered to participate in the experiment and were familiar with Java programming, Maven ecosystem, and API usage. The experience of these participants with Java development is 10+ years. Prior to the experiment, the participants were provided with a 30-minutes tutorial on the tool usage and the experiment process. Each participant were provided with 10 code fragments to perform 10 migration tasks between libraries including *easymock* to *mokito*, and *json* to *gson*. Then, for each of the migration tasks, the developer runs our migration code examples tool that returns a list of examples but exposed to the developers at a random order (at least for our experimental study to avoid biased selection from a ranked list). Then, the developer reviews all the returned examples and picks the top-3 examples that fit her/his preferences and the quality of the examples.

Figure 7 reports the survey results, where the x-axis represents the index of example($k$) in the ranked list, and the y-axis represents the number of times an example@$k$ has been chosen by a developer as their top choice, divided by all choices. In other terms, the y-axis percentage of developers' choice of an example whose rank is $k$. For instance, the value @$k = 1$ is the percentage of how many times the example number one in the ranked list was chosen at the best example.

According to Figure 7, we could see that 59% of developers agreed that the first recommended example is the best example. If we allow the top-2 ranked examples ($k <= 2$), our recommendation already captures 80% of developers' choices, which also improves further to become 94% for top-3 ranked examples ($k <= 3$).

We can conclude that our ranking model efficiently recommends what developers consider to be their decision if they are requested to perform the migration.

> **Summary for RQ3.** The qualitative analysis of our ranking model shows its efficiency to considerably prune the search space for developers when they are searching for good migration examples. Our ranking score was able to match 59% of the developer's chosen examples when recommending top-1 example.

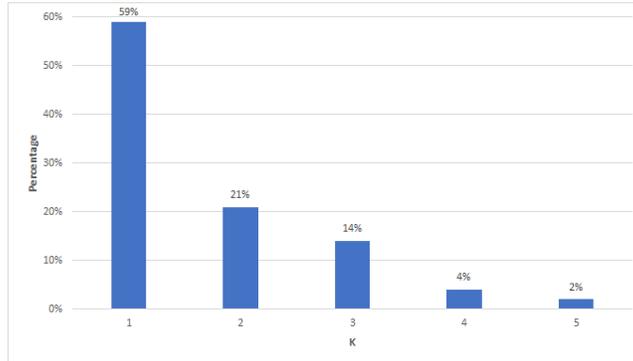

Fig. 7: Percentages of the match between developers choices and the $k^{th}$ example.

## 4 Threats to validity

We report, in this section, potential factors that can threaten the validity of our empirical study.

**Internal Validity.** Our empirical analysis is mainly threatened by the accuracy of the migration dataset. Since our assumption that all studied commits carried at least one migration, any intruding files would be considered as noise to our analysis. We did not perform any rigorous verification concerning the correctness of the dataset, but we did perform various manual checks when gathering the files for statistical analysis and for qualitatively analyze our findings, and we did not notice any single case where the file we were investigating did not contain at least one migration trace.

The second main threat to the validity of our work is the choice of the metrics used in this study. We have chosen coupling, cohesion, complexity, and readability, as being representative to design quality and popular metrics, being used in similar empirical studies [11,12].

The non diverse set of developers, along with the randomness in assigning them the examples, has a direct impact on the results. The choice of experienced and volunteers was to reduce the effect of non interest to the problem resolution. Developers were genuinely interested to support the work, and they were aware of it being potentially published for the community.

**Construct validity.** Threats to construct validity describe concerns about the relationship between theory and observation and, generally, this type of threat is mainly constituted by any errors related to measurements. More precisely, any error in the used tools directly impacts the correctness of our findings. For calculating metrics, we have used popular frameworks and libraries such as Understand. Based on our own humble experience, we did not notice any anomaly while using them.

Moreover, in this study, we did not differentiate between instant and delayed migrations, by combining their results. This may not have allowed to fully un-derstand the difference between both, especially that the instant migration is performed faster than the delayed migration, which may hypothesize that developers may have focused on the correctness of their migrated code, rather than optimizing the design of their system. This remains one of our main future experiments.

**External validity.** Threats to external validity are connected to the generalization of the obtained results. Our empirical study was limited to only open source Java projects. However, we constrained by the tools we use to collect the metrics, and besides Understand, others can only process Java source code. Thus, only the first research question can be extended across languages, if there is such a dataset because the one we have used is also limited to Java libraries.

## 5   Conclusion and Future Work

In this paper, we conducted a large scale empirical study to investigate the impact of software migration between third-party libraries on code quality and comprehension. Our qualitative and empirical analysis indicate that library migrations have a positive impact on software's design, in terms of coupling and cohesion. We also experiment their effect on two state-of-the-art code readability metrics, and we observe an improvement in both metrics. We observed multiple factors that explain the improvement, including the typical better naming conventions and more cohesive API methods. Finally, we leverage structural and readability metrics to define a ranking score for migration examples. To evaluate the effectiveness of our ranking, we surveyed developers to see whether our top recommended examples would match what developers consider to be the best choice. Results show that our top-1 recommended example achieves an agreement of 59%.

These factors drive our future work. We plan on further leveraging API contextual information to recommend better APIs for usage, with respect to a given code fragment. We also plan on extending the structural metrics used to characterize software design quality, such as including the weighted method per class, response for a class, class stability, and depth of inheritance tree.

## References

1. Aggarwal, K.K., Singh, Y., Chhabra, J.K.: An integrated measure of software maintainability. In: Annual Reliability and Maintainability Symposium. 2002 Proceedings (Cat. No. 02CH37318). pp. 235–241. IEEE (2002)
2. Allamanis, M., Sutton, C.: Mining source code repositories at massive scale using language modeling. In: Proceedings of the 10th Working Conference on Mining Software Repositories. pp. 207–216. IEEE Press (2013)
3. Alrubaye, H., Mkaouer, M.W.: Automating the detection of third-party java library migration at the function level. In: Proceedings of the 28th Annual International Conference on Computer Science and Software Engineering. pp. 60–71. IBM Corp. (2018)

4. Alrubaye, H., Mkaouer, M.W., Khokhlov, I., Reznik, L., Ouni, A., Mcgoff, J.: Learning to recommend third-party library migration opportunities at the api level. Applied Soft Computing **90**, 106140 (2020)
5. Alrubaye, H., Mkaouer, M.W., Ouni, A.: Migrationminer: An automated detection tool of third-party java library migration at the method level. In: The International Conference on Software Maintenance and Evolution (ICSME). IEEE Press (2019)
6. Alrubaye, H., Mkaouer, M.W., Ouni, A.: On the use of information retrieval to automate the detection of third-party java library migration at the method level. In: Proceedings of the 27th International Conference on Program Comprehension. pp. 347–357. IEEE Press (2019)
7. Alrubaye, H., Wiem, M.: Variability in library evolution. Software Engineering for Variability Intensive Systems: Foundations and Applications p. 295 (2019)
8. Bavota, G., Linares-Vasquez, M., Bernal-Cardenas, C.E., Di Penta, M., Oliveto, R., Poshyvanyk, D.: The impact of api change-and fault-proneness on the user ratings of android apps. IEEE Transactions on Software Engineering **41**(4), 384–407 (2014)
9. Boehm, B., Basili, V.R.: Software defect reduction top 10 list. Foundations of empirical software engineering: the legacy of Victor R. Basili **426**(37), 426–431 (2005)
10. Buse, R.P., Weimer, W.R.: Learning a metric for code readability. IEEE Transactions on Software Engineering **36**(4), 546–558 (2010)
11. Cedrim, D., Sousa, L., Garcia, A., Gheyi, R.: Does refactoring improve software structural quality? a longitudinal study of 25 projects. In: Proceedings of the 30th Brazilian Symposium on Software Engineering. pp. 73–82. ACM (2016)
12. Chávez, A., Ferreira, I., Fernandes, E., Cedrim, D., Garcia, A.: How does refactoring affect internal quality attributes?: A multi-project study. In: Proceedings of the 31st Brazilian Symposium on Software Engineering. pp. 74–83. ACM (2017)
13. Chidamber, S.R., Kemerer, C.F.: A metrics suite for object oriented design. IEEE Transactions on software engineering **20**(6), 476–493 (1994)
14. Cossette, B.E., Walker, R.J.: Seeking the ground truth: a retroactive study on the evolution and migration of software libraries. In: Proceedings of the ACM SIGSOFT 20th International Symposium on the Foundations of Software Engineering. p. 55. ACM (2012)
15. Endrikat, S., Hanenberg, S., Robbes, R., Stefik, A.: How do api documentation and static typing affect api usability? In: Proceedings of the 36th International Conference on Software Engineering. pp. 632–642. ACM (2014)
16. Kim, M., Nam, J., Yeon, J., Choi, S., Kim, S.: Remi: defect prediction for efficient api testing. In: Proceedings of the 2015 10th Joint Meeting on Foundations of Software Engineering. pp. 990–993. ACM (2015)
17. Kula, R.G., German, D.M., Ouni, A., Ishio, T., Inoue, K.: Do developers update their library dependencies? Empirical Software Engineering **23**(1), 384–417 (2018)
18. Linares-Vásquez, M., Bavota, G., Bernal-Cárdenas, C., Di Penta, M., Oliveto, R., Poshyvanyk, D.: Api change and fault proneness: a threat to the success of android apps. In: Proceedings of the 2013 9th joint meeting on foundations of software engineering. pp. 477–487. ACM (2013)
19. McDonnell, T., Ray, B., Kim, M.: An empirical study of api stability and adoption in the android ecosystem. In: 2013 IEEE International Conference on Software Maintenance. pp. 70–79. IEEE (2013)
20. Pantiuchina, J., Lanza, M., Bavota, G.: Improving code: The (mis) perception of quality metrics. In: 2018 IEEE International Conference on Software Maintenance and Evolution (ICSME). pp. 80–91. IEEE (2018)

21. Romano, D., Raila, P., Pinzger, M., Khomh, F.: Analyzing the impact of antipatterns on change-proneness using fine-grained source code changes. In: 2012 19th Working Conference on Reverse Engineering. pp. 437–446. IEEE (2012)
22. Rugaber, S.: The use of domain knowledge in program understanding. Annals of Software Engineering **9**(1-2), 143–192 (2000)
23. Scalabrino, S., Linares-Vásquez, M., Poshyvanyk, D., Oliveto, R.: Improving code readability models with textual features. In: 2016 IEEE 24th International Conference on Program Comprehension (ICPC). pp. 1–10. IEEE (2016)
24. Shatnawi, A., Seriai, A.D., Sahraoui, H., Alshara, Z.: Reverse engineering reusable software components from object-oriented apis. Journal of Systems and Software **131**, 442–460 (2017)
25. Stevens, W.P., Myers, G.J., Constantine, L.L.: Structured design. IBM Systems Journal **13**(2), 115–139 (1974)
26. Teyton, C., Falleri, J.R., Blanc, X.: Mining library migration graphs. In: Reverse Engineering (WCRE), 2012 19th Working Conference on. pp. 289–298. IEEE (2012)
27. Teyton, C., Falleri, J.R., Blanc, X.: Automatic discovery of function mappings between similar libraries. In: In Reverse Engineering (WCRE), 2013 20th Working Conference on. pp. 192–201. IEEE (2013)
28. Teyton, C., Falleri, J.R., Palyart, M., Blanc, X.: A study of library migrations in java. Journal of Software: Evolution and Process **26**(11), 1030–1052 (2014)